\documentclass[twocolumn,times,tighten]{aastex63}

\received{...}
\revised{...}
\accepted{...}
\submitjournal{ApJL}

\newcommand{\lx}{L_{{\rm x}}}
\newcommand{\ledd}{L_{{\rm Edd}}}

\newcommand{\te}{kT_{\rm e}}
\newcommand{\ec}{E_{\rm c}}
\newcommand{\ergs}{{\rm \,erg\,s^{-1}}}

\graphicspath{{./}{figures/}}

\begin{document}

\shorttitle{Coronal properties of BHXRBs}
\shortauthors{Yan et al.}
\title{Coronal properties of black hole X-ray binaries in the hard state as seen by {\it NuSTAR} and {\it Swift}}

\author[0000-0002-5385-9586]{Zhen Yan}
\affiliation{Key Laboratory for Research in Galaxies and Cosmology, Shanghai Astronomical Observatory, Chinese Academy of Sciences, \\ 80 Nandan Road, Shanghai 200030, China;  zyan@shao.ac.cn, fgxie@shao.ac.cn}

\author[0000-0001-9969-2091]{Fu-Guo Xie}
\affiliation{Key Laboratory for Research in Galaxies and Cosmology, Shanghai Astronomical Observatory, Chinese Academy of Sciences, \\ 80 Nandan Road, Shanghai 200030, China;  zyan@shao.ac.cn, fgxie@shao.ac.cn}

\author[0000-0003-1702-4917]{Wenda Zhang}
\affiliation{Astronomical Institute, Czech Academy of Sciences, Bo\v cn\'i II 1401, CZ-141 00 Prague, Czech Republic;  zhang@asu.cas.cz}



\begin{abstract}

In this work we measure two important phenomenological parameters of corona (and hot accretion flow) in black hole X-ray binaries: the photon index $\Gamma$ and the electron temperature $\te$. Thanks to the capability of {\it NuSTAR} in hard X-rays, we measure these two parameters over six orders of magnitude in the 0.1-100 keV X-ray luminosity $\lx$, from $\sim 5\times10^{38} \ergs$ down to as low as $\sim 5\times10^{32} \ergs$. We confirm the existence of a ``V''-shaped correlation between $\Gamma$ and $\lx$. Surprisingly, we observe a ``$\Lambda$''-shaped correlation between $\te$ and $\lx$. The ``cooler when brighter'' branch in the high luminosity regime ($\lx \ga 3\times 10^{36} \ergs$) agrees with previous results and can be understood under the existing model of Compton scattering in the corona. On the other hand, the apparent ``cooler when fainter'' (positive $\te$-$\lx$ correlation) branch in the low-luminosity regime ($\lx \la 3\times 10^{36} \ergs$) is unexpected, thus it puts a new challenge to existing models of hot accretion flow/corona. 

\end{abstract}


\section{Introduction} \label{sec:intro}
In the current understanding, the hard X-rays from active galactic nuclei (AGN) and black hole X-ray binaries (BHXRB) are produced by a central hot corona (or hot accretion flow; below we use them interchangeably). The soft seed photons, originate either externally from the underlying cold accretion disk or internally from the corona itself, are inversely Compton scattered by the hot electrons in the corona \citep[see reviews by][]{Done2007,Yuan2014}. The corresponding hard X-ray continuum spectrum can be well described by a power-law shape, with a high-energy cutoff at tens to hundreds of keVs. The X-ray luminosity $\lx$, the photon index $\Gamma$ and the cutoff energy $\ec$ determined from X-ray spectral fitting reveal important properties of the corona. For example, the cutoff energy empirically relates to the electron temperature $\te$ as $\ec = 2\sim 3\, \te$ \citep{Petrucci2001}, while the $\Gamma$ is mostly determined by $\te$ and optical depth \citep{Titarchuk1995, Zdziarski1996}. With given $\te$ and $\lx$, the optical depth then provide a constraint on the size of the corona.

Observationally the power-law index $\Gamma$ can be measured much more easily than the electron temperature $\te$. The power-law index can be derived based on the X-ray spectrum below 10 keV, a band covered by most X-ray missions, among which some have high sensitivity. Consequently, the power-law indices have been extensively studied in BHXRBs and AGNs using different samples spanning a large range in X-ray luminosity \citep[e.g.][]{Younes2011, Yang2015}. A ``V" shape in the $\Gamma$-$\lx$ diagram is found: below a certain luminosity ($\sim 1\% \ledd$) the power-law index increases with a decreasing X-ray luminosity (so called ``harder when brighter'') and above that the power-law index increases with an increasing X-ray luminosity (``softer when brighter''). 

The hard and hard-intermediate states of BHXRB, where the X-ray spectrum is usually dominated by the thermal Comptonization of the corona \citep{Done2007,Yuan2014}, which provide an excellent opportunity to study the corona properties. In order to constrain the electron temperature $\te$ (or cutoff energy $\ec$) of the corona, a high-quality X-ray spectrum extending to at least tens of keV are necessary. The most well-studied case is the prototype BHXRB GX 339-4 in its hard and hard-intermediate states, in which an anti-correlation between $E_{\rm c}$ and $\lx$ has been found \citep[e.g.][]{Miyakawa2008,Motta2009}. Similar anti-correlation in hard and/or hard-intermediate states has also been observed in other BHXRBs, such as XTE J1550$-$564 \citep{Rodriguez2003} and GRO J1655$-$40 \citep{Joinet2008}. 

Because of strong background, previous hard X-ray (above 10 keV) instruments are only capable of investigating bright sources, e.g. in most BHXRBs only the $\lx \ga10^{37} \ergs$ regime. Only close sources have good enough hard X-ray spectra to constrain the electron temperature (or cutoff energy). {\it Nuclear Spectroscopic Telescope Array} \citep[{\it NuSTAR,}][]{Harrison2013}, the first focusing telescope in hard X-rays, has unprecedented sensitivity, low background and no pile-up effect. It can provide high signal-to-noise ratio spectra at 3-79 keV energy band for corona at not only high but also low luminosities. 

In this paper, we combine the {\it NuSTAR} and {\it Swift} spectra to study the coronal properties (including $\Gamma$ and $\te$) of the hard state of BHXRBs, with a focus on the low luminosity regime, the electron temperature of which is poorly explored before. Section \ref{sec:data} presents the sample selection and data analysis. Section \ref{sec:results} presents our results. Brief discussions and a summary are devoted to Section \ref{sec:summary}.

\section{Sample selection and Data Analysis}\label{sec:data}

\subsection{Sample/Data Selection}

We list in \autoref{tab:sample} the sources selected in our sample, and \autoref{tab:fit} the details of each observation and spectral results (see Sec. \ref{sec:spec_analysis} for the details of spectral modelling). We assume MAXI J1813$-$095 has a distance of 8 kpc, who lacks distance measurement (see \autoref{tab:sample}).

Our motivation is to investigate the coronal properties, we thus first restrict to BHXRBs that have publicly available {\it NuSTAR} observations. {\it NuSTAR} lacks soft X-ray coverage below $3$ keV, which is crucial for the thermal emission component. We then supplement during the spectral analysis with {\it Swift}/XRT observations that cover the $0.3-10$ keV band. We cross-match the {\it NuSTAR} archive with the {\it Swift} archive for searching quasi-simultaneous observations within one day. The exact observational time of {\it NuSTAR} and {\it Swift} observations is given in \autoref{tab:fit}.

We further select those whose X-ray spectra are dominated by the Comptonization emission. We thus take the following two criteria, one is that the X-ray spectrum can be well fitted with our model ($\chi^2 < 2$), and the other is that the X-ray flux from Comptonization component contributes more than 70\% the total X-ray flux (see details in Section\ \ref{sec:spec_analysis} and \autoref{fig:lratio}). Observations that meet these two criteria are all in the hard or hard-intermediate states \citep{Dunn2010}. During this step, 4U 1630$-$472 and V4641 Sgr are excluded, since none of their observations meet these two criteria.

We notice that there are four {\it NuSTAR} observations that have no quasi-simultaneous (within one day) {\it Swift}/XRT observations. We still include them to enlarge our sample size at low luminosities. Among them, two have {\it Swift} observations within two days: the observation of V404 Cyg on modified Julian date (MJD) 56578 and the observation of GRS 1739$-$278 on MJD 57692. In this work, we still consider them as quasi-simultaneous observations. Meanwhile, the other two, MAXI J1820+070 on MJD 58604 and H1743$-$322 on MJD 57230, have no quasi-simultaneous {\it Swift} observations. They are also included in our sample, but during the spectral modelling, they are fit without the thermal {\it diskbb} component, cf. Sec.\ \ref{sec:spec_analysis}.

We additionally exclude from our sample the following {\it NuSTAR} observations based on various reasons. The observation of H1743$-$322 in its quiescent state on MJD 57250 is not included since it is not robustly detected. The observations of V404 Cyg on MJD 56198 and 57197 associate with many large amplitude flares, during which both the X-ray spectrum and the local absorption are highly variable \citep{Walton2017}. Time-resolved spectra or flux-resolved spectra are necessary \citep[e.g. ][]{Walton2017}. We thus exclude these two observations to avoid the complexities mentioned above.

\begin{table*}
	\begin{center}
	\caption{Sample}
	\label{tab:sample}
	\begin{tabular}{lcccc} 
		\hline
		Source & RA. & Dec. & Distance (kpc) & References \\
		\hline
Swift J1357.2$-$0933 & 13:57:16.84 & -09:32:38.79 & 6.3  &[1] \\
GS 1354$-$64   &  13:58:09.70 & -64:44:05.80 & 25.0  &[2] \\
MAXI J1535$-$571 & 15:35:19.73 &-57:13:48.10  &4.1 & [3] \\ 
GX 339-4 & 17:02:49.31 &-48:47:23.16 & 9.0  &[4]  \\
IGR J17091$-$3624 & 17:09:07.61 & -36:24:25.70 & 12.0 & [5] \\
GRS 1716$-$249 & 17:19:36.93   &-25:01:03.43 & 2.4  & [6] \\
GRS 1739$-$278 & 17:42:40.03 & -27:44:52.70  & 7.5 &  [7]  \\
H1743$-$322   & 17:46:15.60 & -32:14:00.86 & 8.5 & [8] \\
Swift J1753.5$-$0127  &  17:53:28.29 & -01:27:06.26 & 7.15 &  [9] \\
MAXI J1813$-$095 &18:13:34.07  &-09:32:07.30 & &     \\
MAXI J1820+070 &  18:20:21.90 & +07:11:07.30 & 3.46  & [9] \\
GRS 1915+105 & 19:15:11.55 & +10:56:44.76 &  8.6 & [10] \\
Cyg X-1     &  19:58:21.68  & +35:12:05.78 & 1.86   & [11] \\
V404 Cyg & 20:24:03.82 & +33:52:01.90 & 2.4   &  [12] \\
\hline
\end{tabular}
\end{center}
{ References. [1] \citet{ArmasPadilla2014}; [2] \citet{Corral-Santana2016}; [3] \citet{Chauhan2019}; [4] \citet{Heida2017}; [5] \citet{Iyer2015}; [6] \citet{dellaValle1994}; [7] \citet{Yan2017a}; [8] \citet{Steiner2012a}; [9] \citet{Gandhi2019}; [10] \citet{Reid2014}; [11] \citet{Reid2011}; [12] \citet{Miller-Jones2009a}}	   
\end{table*}

\subsection{Data Reduction and Spectral Analysis} \label{sec:spec_analysis}

The {\it NuSTAR} data are processed through the {\scriptsize NUPIPELINE} task of the {\scriptsize NUSTARDAS} package contained in HEASoft 6.25, with calibration files of version 20181030. The source spectra were extracted using a circular region with a radius of 90$\arcsec$ at the source position, and the background spectra were extracted from an annulus with inner and outer radii of 180$\arcsec$ and 200$\arcsec$.

The {\it Swift}/XRT data were firstly processed with {\scriptsize XRTPIPELINE} (v 0.13.4) in order to generate the cleaned event file. We then extract the source and background spectra by using {\scriptsize XSELECT}, while the events at the central pixels were excluded if the data suffer from pile-up effects \citep{Evans2009}.

The joint spectral fitting of quasi-simultaneous {\it Swift}/XRT and {\it NuSTAR} observations were performed using PyXspec with XSPEC 12.10.1. For the modelling, we consider {\it Swift}/XRT and {\it NuSTAR} spectra in $0.5-10$ keV and $4-78$ keV, respectively. 

We fit the spectra with a model that is an absorption of a combination of multi-coloured disk black-body ({\it diskbb}), thermal Comptonization ({\it nthcomp}, \citealt*{Zdziarski1996}) and Gaussian iron line reflection ({\it gauss}) components, i.e. {\it constant*tbabs*(diskbb+nthcomp+gauss)} in {\scriptsize XSPEC} notation. Note that for the two cases without {\it Swift}/XRT observations, {\it diskbb} is omitted. Three observations (on MJD 56578, 56579 and 56628) of V404 Cyg are in the quiescent state, we only consider to use {\it nthcomp} component \citep[see also ][]{Rana2016}. For the absorption by Galactic interstellar medium, the abundances and cross sections are set to \citet{Wilms2000} and \citet{Verner1996}. Note that the column density $N_{\rm H}$ of Swift J1357.2$-$0933, MAXI J1820+070 and GRS 1915+105 is fixed to 0.012$\times 10^{22}$ cm$^{-2}$ \citep{Beri2019}, 0.15$\times 10^{22}$ cm$^{-2}$ \citep{Uttley2018} and 6.5 $\times 10^{22}$ cm$^{-2}$ \citep{Miller2013}, respectively, since they cannot be tightly and consistently constrained under our dataset. 

The thermal Compton model {\it nthcomp} includes three main parameters \citep{Zdziarski1996}: the power-law photon index $\Gamma$, the electron temperature $\te$, and the temperature of seed thermal photons $kT_{\rm s}$. The $kT_{\rm s}$ is tied to the disk temperature when the {\it diskbb} component is applied. During the modelling, the value of {\it constant} is fixed for {\it NuSTAR}/FPMA, but is set free for {\it Swift}/XRT and {\it NuSTAR}/FPMB. Once the best-fitting result is derived, we then use the convolution model {\it cflux} to estimate the 0.1-100 keV X-ray fluxes for the three different components. In most cases parameters $\Gamma$ and $\te$ are well-constrained. Only for the observation of MAXI J1820+070 on MJD 58404, we have a lower-limit constraint on $\te$.

\section{Results} \label{sec:results}

\begin{figure}
    \centering
	\includegraphics[width=\linewidth]{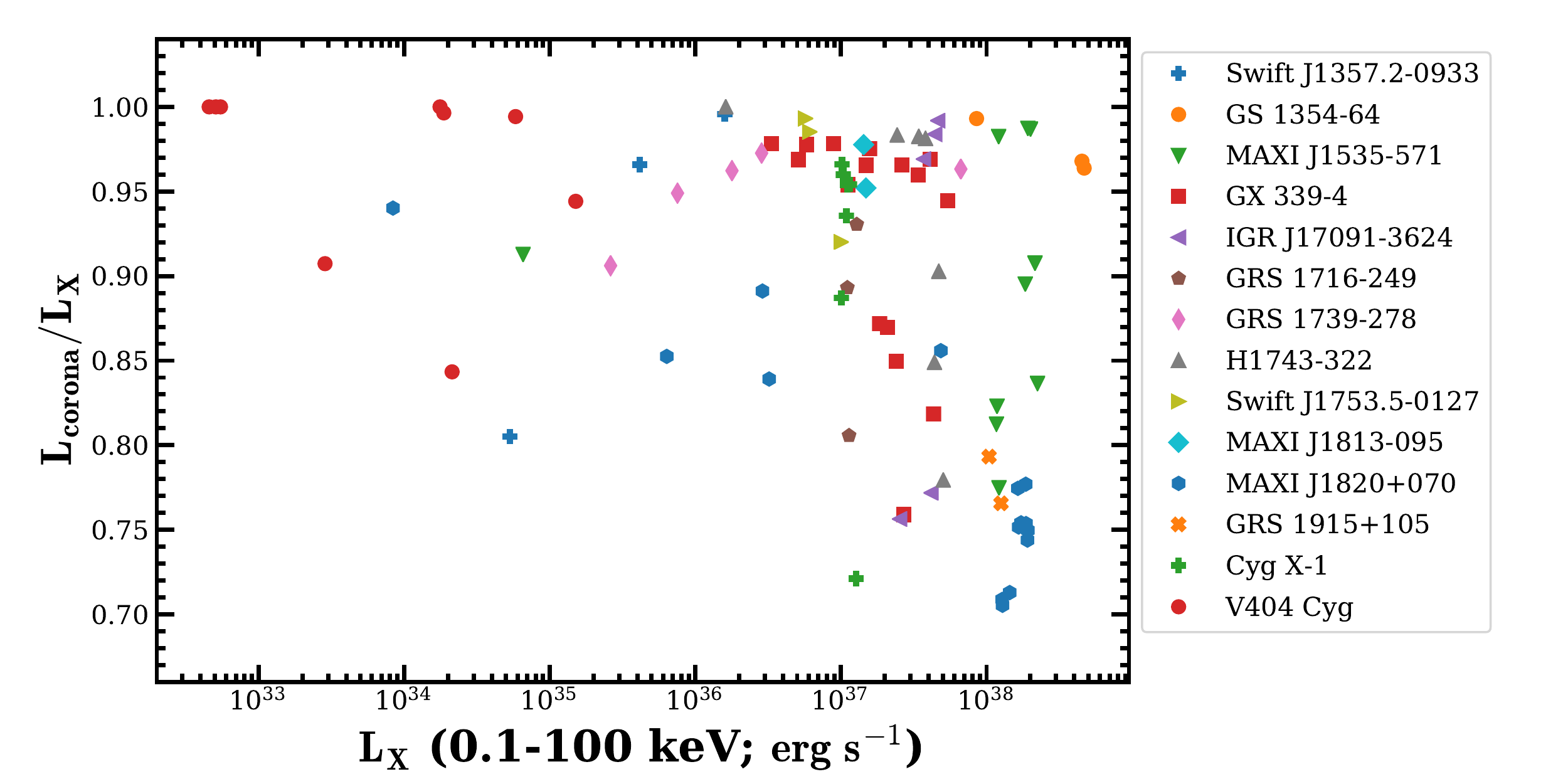}
    \caption{The corona radiation fraction $L_{\rm corona}/\lx$ for BHXRBs in their hard/hard-intermediate state. The uncertainties are not shown here for clarity.}
    \label{fig:lratio}
\end{figure}

\begin{figure*}
    \centering
	\hspace{-0.3cm}\includegraphics[width=0.5\linewidth]{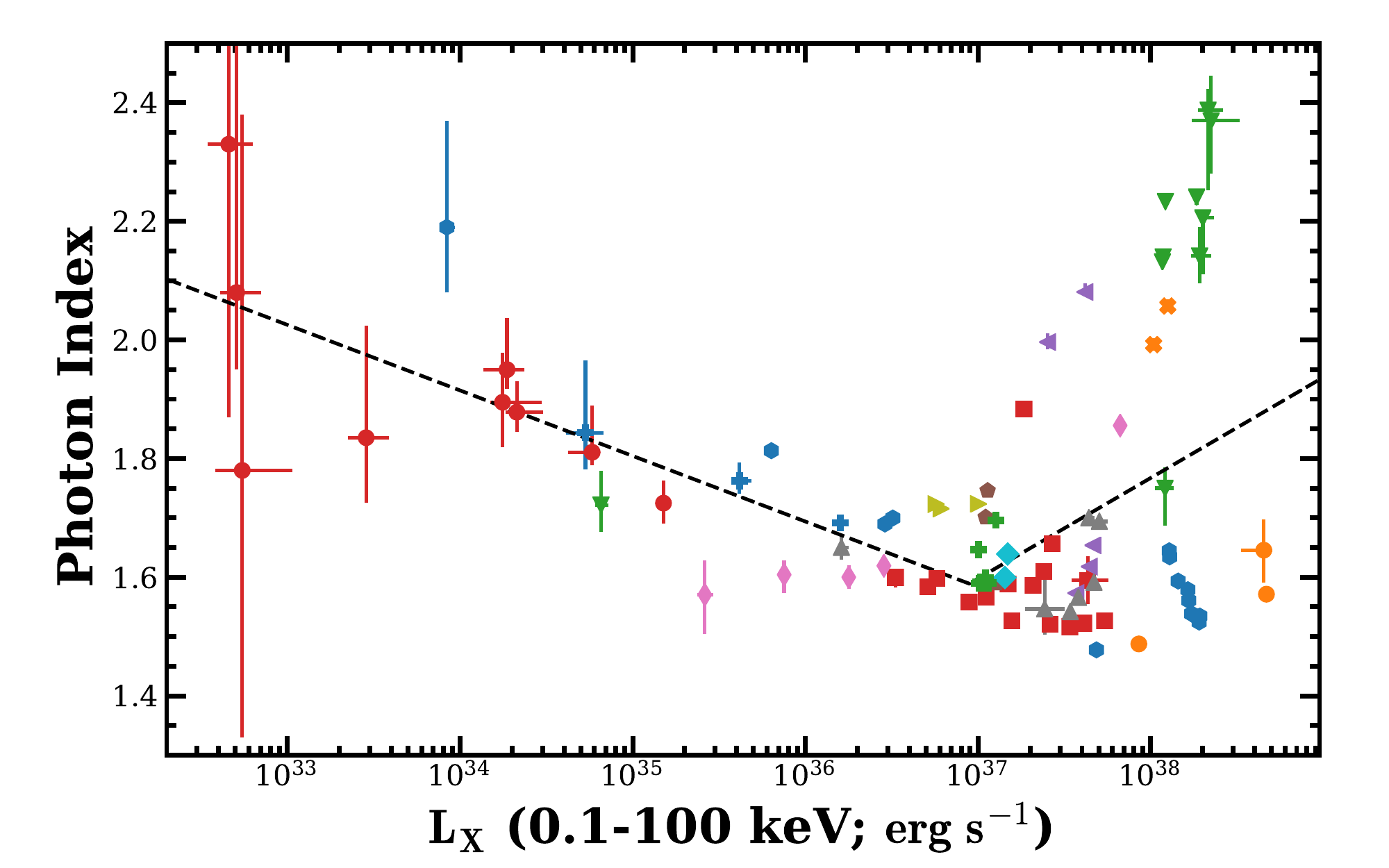}
	\includegraphics[width=0.5\linewidth]{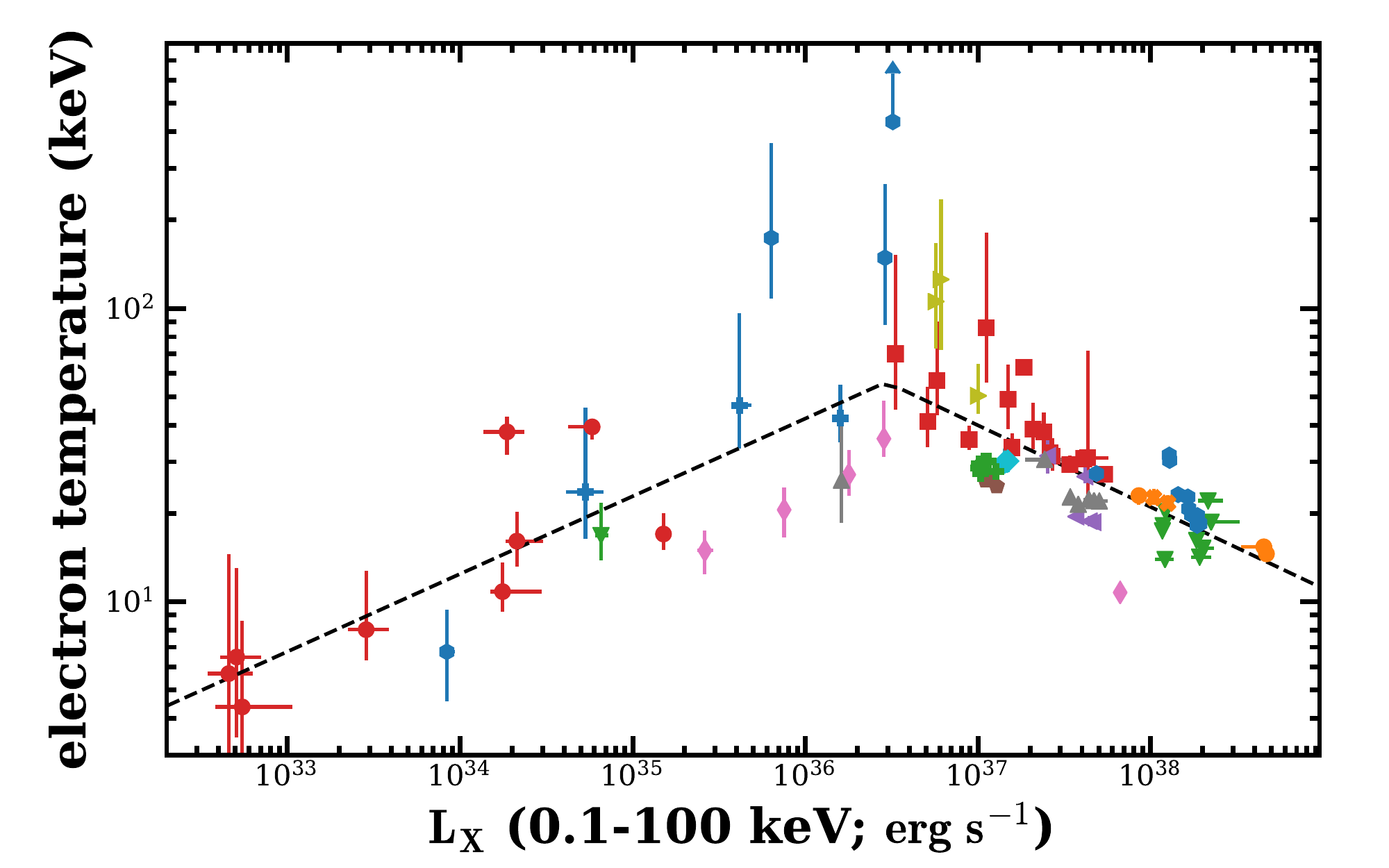}
    \caption{The relationship between photon index $\Gamma$ and the X-ray luminosity $\lx$ ({\it Left Panel}) and that between the X-ray photon index $\Gamma$ and the X-ray luminosity ({\it Right Panel}) for BHXRBs in their hard/hard-intermediate state. The figure legend is the same as \autoref{fig:lratio}.}
    \label{fig:Gamma_te}
\end{figure*}

We first show in \autoref{fig:lratio} the fraction of corona luminosity, which is defined as the {\it nthcomp}-to-all luminosity ratio $L_{\rm corona}/\lx$. Obviously because of the sample selection, all observations have $L_{\rm corona}/\lx > 70\%$, i.e. all are dominated by corona emission. 

The left panel of \autoref{fig:Gamma_te} shows the correlation between the photon index $\Gamma$ and the X-ray luminosity $L_{\rm X}$.
Obviously $\Gamma$ and $\log(L_{\rm X})$ exhibits a ``V"-shaped correlation, consistent with previous results \citep[e.g.,][]{Yamaoka2005, Yuan2007, Wu2008a, Yang2015}. We take the following piecewise linear function to fit the data, 

\begin{equation}
  Y(X) = \left\{
  \begin{array}{cc}
    f_1 (X-X_{c}) + Y_{c}  & \quad (X\leq X_{c}),  \\
    & \\
    f_2 (X-X_{c}) + Y_{c}  & \quad (X\geq X_{c}).
  \end{array} \right.
  \label{eq1}
\end{equation}
By replacing the $X$ and $Y$ with $\log \lx$ and $\Gamma$ in \autoref{eq1}, we fit the $\Gamma$--$\lx$ correlation by using the Levenberg-Marquardt method with the Python package LMFIT \citep{NewvilleMatthew2014}. The best-fit result is shown by the dashed curve in the left panel of \autoref{fig:Gamma_te}. We find the correlation slope turns between positive ($f_{2}=0.17\pm0.05$) and negative ($f_{1}=-0.11\pm0.03$) at $\log \lx = 36.95\pm 0.27$ ($\lx\sim 9\times10^{36} \ergs$). For a 10 $M_{\odot}$ black hole, the turnover happens at nearly 0.01 $\ledd$, which roughly agrees with previous results \citep[e.g.][]{Yuan2007, Wu2008a,  Yang2015}. The Spearman coefficients of the correlation for the data below and above this turning luminosity are $-0.81$ and $0.11$ at the significances of 5.47$\sigma$ and $0.81\sigma$, respectively. Note that it seems that different sources/outbursts follow different tracks of the positive $\Gamma$--$\lx$ correlation branch \citep[see also in ][]{Yamaoka2005, Yang2015}, which results in large scatters and low significance of the positive correlation.

We then investigate the relationship between the electron temperature $\te$ and the X-ray luminosity $\lx$. As shown in the right panel of \autoref{fig:Gamma_te}, they follow a ``$\Lambda$''-shaped $\te$--$\lx$ correlation. Discarding the lower limit value of $\te$ during the fitting, we find the turnover happens at $\log \lx=36.47\pm0.15$ ($\lx\sim 3\times10^{36} \ergs$), which is $\sim 1/3$ that of $\Gamma$--$\lx$ correlation. Considering the uncertainties of the fitting results and large scatter of data, we caution that the exact difference in turnover luminosity awaits future investigations, either statistically or individually. The best-fit slopes of positive and negative branches are $f_{1}=0.26\pm0.03$ and $f_{2}=-0.27\pm0.04$, respectively. The Spearman coefficients are 0.77 and -0.77 at the significances of 4.21$\sigma$ and $8.06\sigma$ for the low and high luminosity branches, respectively. Admittedly some observations with high luminosities are not well-fitted under our model (see \autoref{tab:fit}). The estimation of $\te$ is probably affected by the relativistic reflection components (see discussion in \autoref{sec:refl}). To avoid their contamination, we additionally exclude those data with reduced $\chi^2$ greater than 1.3 and do the fitting again. The slope of the bright branch is now revised to $f_{2}=-0.31\pm0.05$, which agrees with the previous value ($f_{2}=-0.27\pm 0.04$) within uncertainties. The Spearman coefficient of the remaining data is $-0.81$ at a significance of 7.97$\sigma$.

We emphasis that the ``cooler when brighter'' behaviour above $3\times10^{36} \ergs$ is consistent with previous results \citep[e.g. ][]{Miyakawa2008, Motta2009}. However, the ``cooler when fainter'' behaviour with $\lx$ spanning over $\sim $4 orders of magnitude below $3\times10^{36} \ergs$, has never been reported before. 

Due to limitation in the sensitivity of the hard X-ray telescopes, previous studies of electron temperature $\te$ mostly covered the $\lx > 10^{37}\ergs$ high luminosity regime \citep[e.g. ][]{Yamaoka2005, Joinet2008, Miyakawa2008, Motta2009}, among which only a negative $\te$--$\lx$ (in some cases, $\ec$--$\lx$) correlation is reported. For example, \citet{Yamaoka2005} has analyzed the X-ray spectra of 9 BHXRBs with both {\it RXTE} and {\it Beppo-SAX} observations, and found a negative $\ec$--$\lx$ correlation is observed for $\lx > 2\times10^{37} \ergs$. However, as shown in their Figure 3, data of two nearby BHXRBs (XTE J118$+$480 and XTE J1650$-$500) below $2\times10^{36} \ergs$  also hint on a positive relationship. Such turnover is ignored by \citet{Yamaoka2005}, mostly because of the poor $\ec$ measurement quality for $\lx$ in the range $10^{36}$ -- $10^{37}$ $\ergs$.

For completeness, we also estimate the optical depth $\tau$ of the corona according to the following equation \citep{Zdziarski1996}:
\begin{equation}
    \tau\approx\sqrt{\frac{9}{4}+\frac{m_{e}c^{2}}{kT_{e}}\frac{3}{(\Gamma-1)(\Gamma+2)}}-\frac{3}{2},
\end{equation}
i.e., $\tau$ can be crudely determined by $\te$ and $\Gamma$. The results are listed in \autoref{tab:fit}.

\section{Summary and Discussions}\label{sec:summary}

Corona or hot accretion flow, which is responsible for the continuum emission in hard X-rays, is one of the key ingredients in black hole accretion systems. In this work, through a detailed modeling of quasi-simultaneous {\it NuSTAR} + {\it Swift} observations, we analyze the coronal properties of BHXRBs in their hard and hard-intermediate states which spans six orders of magnitudes of luminosity range, from $\sim 5\times10^{32} \ergs$ to $\sim 5\times10^{38} \ergs$. 

We confirm previous work that $\Gamma$ and $\lx$ follows a ``V''-shaped correlation, and the turnover happens at $\lx\approx 9\times10^{36}\ergs$. Meanwhile, we find unexpectedly a ``$\Lambda$''-shaped relationship between electron temperature $\te$ and $\lx$, i.e. it shows a ``cooler when brighter'' behavior when $\lx \ga 3\times10^{36}\ergs$, and an opposite ``cooler when fainter'' behavior when $\lx \la 3\times10^{36}\ergs$. As discussed below, this result challenges the existing models of BHXRBs in hard state.

\subsection{Theoretical Implication}
The leading model for the hard and hard-intermediate states of BHXRBs is the truncated accretion--jet model \citep{Done2007, Yuan2014}. In this model, the hard X-rays are produced by the inverse Compton scattering of electrons within the inner hot accretion flow, the residual thermal emission below $\sim1-2$ keV is mostly produced by the outer truncated cold Sukura-Suyaev disk (SSD), and the radio up to infrared is from a relativistic jet.

This model has been applied to understand the ``V''-shaped $\Gamma-\lx$ correlation, where the primary reason for the opposite spectral behavior relates to the change in the origin/source of seed photons for the Compton scattering process \citep{Yang2015}. At the relative high accretion rate, the seed photons are mainly external, from quasi-thermal emission of cold clumps within the hot accretion flow \citep{Yang2015} and/or SSD \citep{Qiao2013}, while at low accretion rate they are mainly internal, from synchrotron emission within hot accretion flow/corona itself.

Despite the change in the seed photons, we should {\it always} expect that more radiative cooling (i.e. higher $\lx$) will result in cooler electrons, i.e. it should always follow a ``cooler when brighter'' track in $\te-\lx$ correlation, the case as observed for $\lx > 3\times10^{36}\ergs$. $\lx < 3\times10^{36}\ergs$ represents the typical regime of the hot accretion flow, where theory predicts that the hot accretion flow will be more close to the ideal non-radiative or at least radiatively-inefficient case \citep{Yuan2014}. In this regime, the electron temperature will reach to its maximal value, nearly independent of accretion rate (or $\lx$). Obviously, such theoretical expectation contradicts to the observation of the positive $\te-\lx$ correlation branch, where the electrons become cooler at lower luminosity. Interestingly, we notice that \citet{Yu2015b} investigated an optically-thin but radiatively-efficient accretion flow, where strong magnetic fields are considered to avoid thermal instabilities. In this model, a ``cooler when fainter'' behavior is indeed achieved \citep{Yu2015b}. On the other hand, it has been noticed for a long time that quantitatively the observed electron temperature $\te$ is systematically lower than that predicted by hot accretion flow \citep[see e.g. ][]{Yuan2004,Xie2010}. Our measurements of $\te$ are also much below the prediction of hot accretion flow, especially at the low luminosity regime.

The other scenario in literature considers the possibility of electron-positron pair production \citep{Svensson1984,Coppi1999}. In this model, electron-positron pairs are created when energetic photons collide with one another. This process acts as an $\lx/R$-dependent thermostat \citep[R is the size of emission site. See ][]{Svensson1984,Coppi1999}, thus plays a major role in determining the outgoing spectrum and overall composition of the corona. This model predicts a negative $\te-\lx/R$ correlation \citep[see e.g.,][]{Fabian2015}. Obviously our results at low luminosities also contradict to this model.

Another competing scenario suggested in the literature for the hard state of BHXRBs is the maximal jet model \citep[e.g.,][]{Markoff2001, Markoff2005}.\footnote{Because of adopting an incorrect Bernoulli equation, the existence of the maximal jet model is challenged \citep{Zdziarski2016}.} In this model, the X-rays are the synchrotron emission of the accelerated particles in the jet base.  However, synchrotron emission cannot produce a sharp cut-off feature in hard X-rays \citep{Zdziarski2003}. Moreover, it is still unclear to us how to understand in this model not only the newly discovered ``$\Lambda$''-shaped $\te-\lx$ relationship, but also the well-established ``V''-shaped $\Gamma-\lx$ correlation.

\subsection{Reflection Emission}\label{sec:refl}

The reflection emission is known to have affects on the determination of $\te$ \citep{Garcia2015, Fabian2015}. However, the reflection emission in the low luminosity regime ($\lx \la 10^{37}\ergs$) is systematically weak \citep[e.g. ][]{Furst2016a,Beri2019}. For example, the reflection fraction is found to be less than $5\%$ for a joint {\it NuSTAR} and {\it XMM-Newton} observation of GRS 1739$-$278 at $\lx \sim (2-3)\times10^{35}\ergs$ \citep{Furst2016a}. Consequently the cut-off energy and the $\te$ constrained from Comptonization model is consistent with those derived by reflection models \citep{Furst2016a}. Since we expect the reflection fraction to decrease with decreasing $\lx$, we thus argue that for $\lx \la 3\times10^{36}\ergs$ the estimation of $\te$ is insensitive to whichever reflection model adopted. We also emphasis that GRS 1739$-$278 individually also follows a positive $\te$--$\lx$ correlation at low luminosity regime, as being clearly demonstrated in \autoref{fig:Gamma_te}. 

We also examine our results by using the non-relativistic reflection model {\it xillverCp}, which includes {\it nthcomp} as an incident spectrum. We find that (not shown here), although in numerous cases the exact value of $\te$ is different from that derived by {\it nthcomp}, we still obtain  a ``$\Lambda$'' shape $\te$--$\lx$ correlation. The best-fit values of slopes of positive and negative branches are $f_{1}=0.11\pm0.02$ and  $f_{2}=-0.19\pm0.06$, respectively. The Spearman coefficients for the low and high luminosity branches are respectively 0.88 and -0.36 at the significances of 6.08$\sigma$ and $2.10\sigma$. The separation of two branches locates at the luminosity $\log L_{X}=36.94\pm0.24$. So our main conclusions are still solid with the reflection model. We do not report our results based on {\it xillverCp} model, mainly because some bright observations are poorly fitted by this model (with reduced chi-square larger than two). The relativistic reflection model may be required for accurate estimation of $\te$ in those bright observations. For example, \citet{Basak2017} also analysed the {\it NuSTAR} data of the bright persistent BHXRB Cyg X-1. The $\te$ derived by sophisticated models (including multiple reflection components) is about 90 keV, which is roughly three times larger than our results \citep[see also ][]{Ibragimov2005}. On the other hand, this negative correlation has been demonstrated in previous studies with different models \citep[e.g. ][]{Yamaoka2005,Fabian2015}.   

In this work we are not aimed at deriving the exact value of $\te$, but instead at the trend of the $\te$--$\lx$ correlation, especially at low luminosity regime, which has been poorly investigated before. For this motivation, we adopt the same model for all the observations (see Section \ref{sec:spec_analysis}) instead of elaborately examining/modelling each observation in detail. The discrepancy of $\te$ in different models is beyond the scope of this paper.

\subsection{Quiescent state of BHXRBs}
All the data points with $\lx<10^{33}\ergs$ are from V404 Cyg in its quiescent state \citep{Plotkin2017a}. A component with a cut-off at $\sim 20$ keV is clearly detected \citep{Rana2016}. However, the origin of the hard X-ray emission in the quiescent state remains unclear, and various models have been proposed \citep[e.g. ][]{Narayan1996, Xie2014,Rana2016,Plotkin2017a}.

Although we cannot directly eliminate the debate, the data points in quiescent state roughly agree with the extrapolation (to fainter end) of the two correlations observed in the low-luminosity regime of hard state, i.e., the negative $\Gamma-\lx$ relationship  \citep[see left panel of \autoref{fig:Gamma_te} and also][]{Plotkin2017a} and the positive $\te-\lx$ one (see right panel of \autoref{fig:Gamma_te}). This implies that the hard X-rays in the quiescent state has the same origin as that in the faint hard state, although the accretion physics in the latter is also unclear yet.

\subsection{Corona in AGNs}

A similar ``V''-shaped $\Gamma-\lx$ correlation has also been observed in active galactic nuclei \citep[AGN; e.g.][]{Younes2011,Yang2015}, implying that the accretion physics in both stellar and super-massive black hole systems are at least similar \citep{Yang2015}.

The electron temperature of corona in AGNs have been probed extensively by various hard X-ray missions \citep[e.g.][]{Dadina2007,Molina2013, Tortosa2018}. A negative correlation between electron temperature (or cut-off energy) and the luminosity in the Eddington unit is reported by a sample of AGNs as observed by {\it Swift}/BAT \citep[e.g.][]{Ricci2018}. However, this result is not confirmed by a sample that although smaller in size but better in data quality \citep{Molina2019,Rani2019}. A large sample of AGN (especially including the low luminosity ones) is necessary to reach a consensus of coronal properties. 
If the accretion physics is similar, we will expect to observe a similar $\Lambda$-shaped correlation between $\te$ and $\lx$ in AGNs; or in other words, a different result will reveal distinctive differences between AGNs and BHXRBs.

\acknowledgments
We would like to thank Wenfei Yu and Anabella Araudo for discussion and Andrzej Zdziarski for his helpful comments. This research has made use of data and software provided by the High Energy Astrophysics Science Archive Research Center (HEASARC), which is a service of the Astrophysics Science Division at NASA/GSFC and the High Energy Astrophysics Division of the Smithsonian Astrophysical Observatory. Z.Y. and F.G.X. are supported in part by the National Program on Key Research and Development Project of China (grant 2016YFA0400704) and the Natural Science Foundation of China (grants 11773055, 11873074, U1938114 and U1838203). W.Z. acknowledges financial support provided by Czech Science Foundation grant 17-02430S.
W.Z. is also supported by the project RVO:67985815.

\vspace{5mm}
\facilities{NuSTAR, Swift(XRT)}


\clearpage
\appendix
\section{Appendix Table}

\begin{longrotatetable}
\begin{deluxetable*}{lrcrcccccc}
 \centerwidetable
\tablecaption{List of quasi-simultaneous observations of {\it Swift} and {\it NuSTAR} and best-fitting parameters\label{tab:fit}}
\tabletypesize{\scriptsize}
\tablehead{
\colhead{Source} & \multicolumn{2}{c}{{\it NuSTAR}}  & 
\multicolumn{2}{c}{{\it Swift}} & \colhead{$\Gamma$} &\colhead{$kT_{e}$} & $\tau$& \colhead{Flux (0.1-100\,keV)}  & \colhead{$\chi^{2}/dof$}\\ 
\cline{2-5}
\colhead{} & \colhead{ObsID} & \colhead{Date(MJD)} & \colhead{ObsID} & 
\colhead{Date (MJD)} & \colhead{} & \colhead{ (keV) } &\colhead{}  &\colhead{(erg s$^{-1}$ cm$^{-2}$)}& \colhead{} 
} 
\startdata
Swift J1357.2-0933&90201057002&57871.54&00088094002&57871.62&$1.69_{-0.01}^{+0.01}$&$42.21_{-7.33}^{+12.84}$&$2.56_{-0.42}^{+0.42}$&$3.70_{-0.07}^{+0.19} \times 10^{-10}$&$1160.47/1210$ \\
Swift J1357.2-0933&90301005002&57914.57&00031918066&57914.61&$1.76_{-0.02}^{+0.03}$&$46.70_{-13.39}^{+49.76}$&$2.20_{-1.05}^{+1.05}$&$9.65_{-0.62}^{+1.70} \times 10^{-11}$&$385.48/422$ \\
Swift J1357.2-0933&90501325002&58632.51&00031918085&58632.29&$1.84_{-0.06}^{+0.12}$&$23.68_{-7.29}^{+22.26}$&$3.21_{-1.35}^{+1.35}$&$1.24_{-0.29}^{+0.34} \times 10^{-11}$&$218.04/209$ \\
\hline
GS 1354-64&90101006002&57186.29&00033811005&57186.61&$1.49_{-0.00}^{+0.01}$&$23.00_{-1.56}^{+1.56}$&$4.94_{-0.21}^{+0.21}$&$1.14_{-0.03}^{+0.03} \times 10^{-9}$&$1954.83/1839$ \\
GS 1354-64&90101006004&57214.57&00033811017&57214.35&$1.57_{-0.00}^{+0.00}$&$14.57_{-0.26}^{+0.20}$&$5.84_{-0.06}^{+0.06}$&$6.26_{-0.17}^{+0.12} \times 10^{-9}$&$3301.82/2805$ \\
GS 1354-64&90101006006&57240.30&00033811040&57240.55&$1.65_{-0.05}^{+0.05}$&$15.39_{-0.68}^{+0.54}$&$5.18_{-0.33}^{+0.33}$&$6.04_{-1.60}^{+0.75} \times 10^{-9}$&$3789.37/2852$ \\
\hline
MAXI J1535-571&90301013002&58003.78&00010264003&58004.28&$1.75_{-0.06}^{+0.03}$&$13.95_{-0.49}^{+0.35}$&$4.93_{-0.26}^{+0.26}$&$6.02_{-0.75}^{+0.76} \times 10^{-8}$&$6878.04/3923$ \\
MAXI J1535-571&80302309002&58008.54&00010264005&58008.26&$2.24_{-0.01}^{+0.01}$&$16.21_{-0.68}^{+0.75}$&$3.00_{-0.09}^{+0.09}$&$9.16_{-0.16}^{+0.20} \times 10^{-8}$&$3117.38/2613$ \\
MAXI J1535-571&80402302002&58010.22&00010264006&58010.93&$2.14_{-0.05}^{+0.05}$&$14.21_{-0.32}^{+0.49}$&$3.51_{-0.14}^{+0.14}$&$9.55_{-1.06}^{+1.63} \times 10^{-8}$&$3133.83/2795$ \\
MAXI J1535-571&80402302004&58010.55&00010264006&58010.93&$2.21_{-0.10}^{+0.01}$&$15.23_{-0.24}^{+0.55}$&$3.20_{-0.13}^{+0.13}$&$9.96_{-0.52}^{+1.59} \times 10^{-8}$&$3235.94/2821$ \\
MAXI J1535-571&80302309004&58013.17&00088245001&58013.18&$2.39_{-0.13}^{+0.04}$&$22.11_{-0.78}^{+0.72}$&$2.19_{-0.14}^{+0.14}$&$1.07_{-0.13}^{+0.23} \times 10^{-7}$&$3795.49/2522$ \\
MAXI J1535-571&80302309006&58013.44&00088245001&58013.18&$2.37_{-0.09}^{+0.08}$&$18.69_{-0.57}^{+0.77}$&$2.49_{-0.15}^{+0.15}$&$1.11_{-0.25}^{+0.51} \times 10^{-7}$&$4733.99/2830$ \\
MAXI J1535-571&80302309014&58048.88&00088245004&58048.99&$2.23_{-0.01}^{+0.01}$&$20.33_{-0.81}^{+1.00}$&$2.59_{-0.08}^{+0.08}$&$6.05_{-0.12}^{+0.13} \times 10^{-8}$&$3414.27/2786$ \\
MAXI J1535-571&80402302009&58050.96&00088246001&58050.98&$2.14_{-0.01}^{+0.01}$&$18.19_{-0.73}^{+1.20}$&$2.98_{-0.11}^{+0.11}$&$5.86_{-0.14}^{+0.16} \times 10^{-8}$&$2582.16/2478$ \\
MAXI J1535-571&80402302010&58051.36&00088246001&58050.98&$2.13_{-0.01}^{+0.01}$&$17.47_{-0.61}^{+0.84}$&$3.08_{-0.09}^{+0.09}$&$5.81_{-0.13}^{+0.16} \times 10^{-8}$&$2978.61/2781$ \\
MAXI J1535-571&90501314002&58587.55&00088862002&58587.54&$1.72_{-0.05}^{+0.06}$&$16.84_{-3.02}^{+4.88}$&$4.51_{-0.70}^{+0.70}$&$3.26_{-0.25}^{+0.35} \times 10^{-11}$&$656.60/789$ \\
\hline
GX 339-4&80001013002&56515.99&00032490015&56516.02&$1.53_{-0.00}^{+0.00}$&$33.55_{-3.01}^{+3.98}$&$3.68_{-0.25}^{+0.25}$&$1.62_{-0.02}^{+0.02} \times 10^{-9}$&$2086.36/2213$ \\
GX 339-4&80001013004&56520.71&00080180001&56520.77&$1.52_{-0.00}^{+0.00}$&$32.14_{-3.06}^{+4.20}$&$3.82_{-0.28}^{+0.28}$&$2.70_{-0.05}^{+0.05} \times 10^{-9}$&$2484.27/2453$ \\
GX 339-4&80001013006&56528.53&00080180002&56528.17&$1.52_{-0.00}^{+0.00}$&$30.73_{-2.06}^{+2.36}$&$3.92_{-0.18}^{+0.18}$&$4.22_{-0.05}^{+0.05} \times 10^{-9}$&$2793.80/2706$ \\
GX 339-4&80001013007&56538.38&00032898013&56537.79&$1.59_{-0.04}^{+0.04}$&$30.95_{-7.75}^{+40.85}$&$3.54_{-1.81}^{+1.81}$&$4.46_{-0.87}^{+1.38} \times 10^{-9}$&$577.60/563$ \\
GX 339-4&80001013008&56538.41&00032898013&56537.79&$1.53_{-0.00}^{+0.00}$&$27.22_{-0.97}^{+0.87}$&$4.21_{-0.09}^{+0.09}$&$5.58_{-0.07}^{+0.08} \times 10^{-9}$&$3632.28/3414$ \\
GX 339-4&80001013010&56581.99&00032898035&56581.56&$1.56_{-0.00}^{+0.00}$&$35.67_{-2.69}^{+4.22}$&$3.39_{-0.21}^{+0.21}$&$9.16_{-0.10}^{+0.14} \times 10^{-10}$&$2458.88/2473$ \\
GX 339-4&80102011002&57262.55&00032898124&57263.37&$1.66_{-0.01}^{+0.01}$&$31.38_{-3.43}^{+4.79}$&$3.25_{-0.28}^{+0.28}$&$2.78_{-0.08}^{+0.08} \times 10^{-9}$&$2289.81/2219$ \\
GX 339-4&80102011003&57267.49&00032898126&57268.03&$1.88_{-0.00}^{+0.00}$&$63.06_{-0.01}^{+0.00}$&$1.55_{-0.00}^{+0.00}$&$1.90_{-0.00}^{+0.00} \times 10^{-9}$&$509.07/536$ \\
GX 339-4&80102011004&57267.53&00032898126&57268.03&$1.61_{-0.01}^{+0.00}$&$37.94_{-4.73}^{+6.16}$&$3.04_{-0.29}^{+0.29}$&$2.48_{-0.08}^{+0.09} \times 10^{-9}$&$2268.84/2220$ \\
GX 339-4&80102011006&57272.62&00032898130&57272.02&$1.59_{-0.01}^{+0.01}$&$38.70_{-5.64}^{+8.90}$&$3.09_{-0.39}^{+0.39}$&$2.15_{-0.06}^{+0.08} \times 10^{-9}$&$2137.45/2144$ \\
GX 339-4&80102011008&57277.66&00081534001&57277.68&$1.59_{-0.01}^{+0.01}$&$48.93_{-10.10}^{+15.36}$&$2.63_{-0.47}^{+0.47}$&$1.54_{-0.03}^{+0.03} \times 10^{-9}$&$1821.91/1760$ \\
GX 339-4&80102011010&57282.42&00032898138&57282.00&$1.57_{-0.01}^{+0.01}$&$85.72_{-29.90}^{+95.55}$&$1.83_{-0.97}^{+0.97}$&$1.15_{-0.03}^{+0.03} \times 10^{-9}$&$1992.58/2004$ \\
GX 339-4&80102011012&57295.05&00081534005&57295.23&$1.60_{-0.01}^{+0.01}$&$56.66_{-13.38}^{+33.73}$&$2.35_{-0.68}^{+0.68}$&$5.97_{-0.12}^{+0.16} \times 10^{-10}$&$1539.28/1547$ \\
GX 339-4&80302304002&58028.15&00032898148&58027.28&$1.60_{-0.02}^{+0.02}$&$70.01_{-24.91}^{+81.88}$&$2.02_{-1.10}^{+1.10}$&$3.43_{-0.10}^{+0.12} \times 10^{-10}$&$1029.14/982$ \\
GX 339-4&80302304004&58051.57&00032898158&58051.38&$1.52_{-0.01}^{+0.00}$&$29.39_{-1.96}^{+2.14}$&$4.06_{-0.18}^{+0.18}$&$3.51_{-0.06}^{+0.09} \times 10^{-9}$&$2379.42/2302$ \\
GX 339-4&80302304007&58148.36&00032898163&58148.16&$1.58_{-0.01}^{+0.01}$&$41.17_{-7.61}^{+12.90}$&$2.98_{-0.50}^{+0.50}$&$5.28_{-0.15}^{+0.13} \times 10^{-10}$&$1311.36/1337$ \\
\hline
IGR J17091-3624&80001041002&57454.08&00031921099&57454.08&$1.57_{-0.00}^{+0.00}$&$19.52_{-0.54}^{+0.95}$&$4.87_{-0.12}^{+0.12}$&$2.14_{-0.03}^{+0.07} \times 10^{-9}$&$2760.57/2665$ \\
IGR J17091-3624&80202014002&57459.59&00031921104&57459.58&$1.62_{-0.00}^{+0.01}$&$18.83_{-0.71}^{+0.82}$&$4.72_{-0.12}^{+0.12}$&$2.56_{-0.04}^{+0.04} \times 10^{-9}$&$2386.88/2381$ \\
IGR J17091-3624&80202014004&57461.81&00031921106&57461.90&$1.65_{-0.00}^{+0.00}$&$18.67_{-0.64}^{+0.66}$&$4.55_{-0.10}^{+0.10}$&$2.68_{-0.04}^{+0.04} \times 10^{-9}$&$2332.66/2283$ \\
IGR J17091-3624&80202014006&57476.11&00031921118&57475.59&$2.08_{-0.01}^{+0.02}$&$26.72_{-1.35}^{+3.86}$&$2.41_{-0.16}^{+0.16}$&$2.41_{-0.09}^{+0.09} \times 10^{-9}$&$2354.90/2183$ \\
IGR J17091-3624&80202015004&57534.66&00081917002&57534.67&$2.00_{-0.01}^{+0.02}$&$31.41_{-4.07}^{+4.91}$&$2.31_{-0.23}^{+0.23}$&$1.47_{-0.06}^{+0.07} \times 10^{-9}$&$1951.30/1943$ \\
\hline
GRS 1716-249&80201034007&57781.31&00034924001&57781.00&$1.59_{-0.00}^{+0.00}$&$24.77_{-0.44}^{+0.51}$&$4.11_{-0.05}^{+0.05}$&$1.86_{-0.01}^{+0.03} \times 10^{-8}$&$4428.11/4079$ \\
GRS 1716-249&90202055002&57850.60&00034924029&57850.37&$1.70_{-0.00}^{+0.00}$&$25.95_{-0.84}^{+1.21}$&$3.50_{-0.09}^{+0.09}$&$1.60_{-0.02}^{+0.02} \times 10^{-8}$&$4352.30/3511$ \\
GRS 1716-249&90202055004&57853.69&00034924031&57853.88&$1.75_{-0.00}^{+0.00}$&$26.78_{-1.36}^{+1.27}$&$3.27_{-0.11}^{+0.11}$&$1.65_{-0.02}^{+0.02} \times 10^{-8}$&$3588.63/3340$ \\
\hline
GRS 1739-278&80002018002&56742.67&00033203003&56742.15&$1.86_{-0.00}^{+0.01}$&$10.75_{-0.10}^{+0.12}$&$5.24_{-0.04}^{+0.04}$&$9.89_{-0.11}^{+0.12} \times 10^{-9}$&$4436.78/3162$ \\
GRS 1739-278&80101050002&57280.88&00081764002&57280.93&$1.57_{-0.07}^{+0.06}$&$14.95_{-2.51}^{+2.56}$&$5.75_{-0.73}^{+0.73}$&$3.89_{-0.36}^{+0.44} \times 10^{-11}$&$767.09/793$ \\
GRS 1739-278&80102101002&57660.89&00033812058&57661.07&$1.62_{-0.01}^{+0.02}$&$35.95_{-4.77}^{+12.50}$&$3.11_{-0.50}^{+0.50}$&$4.23_{-0.11}^{+0.16} \times 10^{-10}$&$1268.15/1339$ \\
GRS 1739-278&80102101004&57680.63&00081979001&57680.76&$1.60_{-0.03}^{+0.02}$&$20.57_{-4.00}^{+3.90}$&$4.54_{-0.56}^{+0.56}$&$1.12_{-0.07}^{+0.07} \times 10^{-10}$&$847.17/1002$ \\
GRS 1739-278&80102101005&57692.84&00033812067&57691.10&$1.60_{-0.02}^{+0.02}$&$27.13_{-4.17}^{+5.83}$&$3.83_{-0.46}^{+0.46}$&$2.65_{-0.10}^{+0.11} \times 10^{-10}$&$1033.37/1026$\\
\hline
H1743-322&80001044002&56918.65&00031121055&56918.35&$1.54_{-0.00}^{+0.00}$&$22.70_{-0.61}^{+0.84}$&$4.62_{-0.09}^{+0.09}$&$3.96_{-0.03}^{+0.05} \times 10^{-9}$&$3246.68/3143$ \\
H1743-322&80001044004&56923.76&00031121061&56924.08&$1.59_{-0.01}^{+0.00}$&$22.08_{-0.64}^{+0.60}$&$4.41_{-0.08}^{+0.08}$&$5.43_{-0.12}^{+0.15} \times 10^{-9}$&$3341.25/3185$ \\
H1743-322&80001044006&56939.76&00080797002&56940.20&$1.57_{-0.00}^{+0.00}$&$21.45_{-0.86}^{+0.66}$&$4.64_{-0.10}^{+0.10}$&$4.40_{-0.06}^{+0.05} \times 10^{-9}$&$2901.29/2924$ \\
H1743-322&80002040002&57206.12&00080797003&57206.31&$1.55_{-0.04}^{+0.07}$&$30.50_{-1.71}^{+1.18}$&$3.81_{-0.31}^{+0.31}$&$2.81_{-0.65}^{+0.86} \times 10^{-9}$&$2338.93/2357$ \\
H1743-322&80002040004&57230.16&\nodata&\nodata&$1.65_{-0.02}^{+0.02}$&$25.92_{-7.34}^{+18.68}$&$3.71_{-1.20}^{+1.20}$&$1.87_{-0.10}^{+0.20} \times 10^{-10}$&$931.74/959$\\
H1743-322&80202012002&57460.07&00031441056&57460.24&$1.69_{-0.00}^{+0.00}$&$22.02_{-0.59}^{+0.79}$&$3.92_{-0.08}^{+0.08}$&$5.83_{-0.61}^{+0.71} \times 10^{-9}$&$3504.28/3303$ \\
H1743-322&80202012004&57462.28&00031441058&57462.90&$1.70_{-0.00}^{+0.00}$&$22.33_{-0.47}^{+0.74}$&$3.86_{-0.07}^{+0.07}$&$5.07_{-0.34}^{+0.40} \times 10^{-9}$&$3495.04/3386$ \\
\hline
Swift J1753.5-0127&80002021002&56751.15&00080730001&56752.02&$1.72_{-0.00}^{+0.00}$&$105.54_{-32.43}^{+61.31}$&$1.26_{-0.43}^{+0.43}$&$9.33_{-0.05}^{+0.17} \times 10^{-10}$&$1903.75/1917$\\
Swift J1753.5-0127&80002021003&56751.89&00080730001&56752.02&$1.72_{-0.01}^{+0.00}$&$125.46_{-53.18}^{+110.42}$&$1.12_{-0.57}^{+0.57}$&$9.98_{-0.18}^{+0.20} \times 10^{-10}$&$1908.30/1911$\\
Swift J1753.5-0127&30001148002&56913.41&00080770001&56913.49&$1.72_{-0.00}^{+0.00}$&$50.34_{-6.74}^{+14.40}$&$2.18_{-0.32}^{+0.32}$&$1.64_{-0.04}^{+0.04} \times 10^{-9}$&$2400.69/2316$\\
\hline
MAXI J1813-095&80402303004&58183.60&00088654002&58183.56&$1.60_{-0.01}^{+0.01}$&$29.94_{-2.49}^{+3.26}$&$3.60_{-0.22}^{+0.22}$&$1.87_{-0.03}^{+0.03} \times 10^{-9}$&$2084.45/2127$\\
MAXI J1813-095&80402303006&58202.73&00088654004&58202.84&$1.64_{-0.01}^{+0.01}$&$30.15_{-2.45}^{+3.41}$&$3.41_{-0.22}^{+0.22}$&$1.94_{-0.05}^{+0.07} \times 10^{-9}$&$2199.20/2162$\\
MAXI J1820+070&90401309002&58191.85&00010627001&58191.87&$1.48_{-0.00}^{+0.00}$&$27.32_{-0.61}^{+0.59}$&$4.50_{-0.06}^{+0.06}$&$3.38_{-0.05}^{+0.05} \times 10^{-8}$&$4746.61/3842$\\
MAXI J1820+070&90401309004&58198.02&00010627009&58197.77&$1.53_{-0.00}^{+0.00}$&$19.65_{-0.22}^{+0.53}$&$5.11_{-0.06}^{+0.06}$&$1.30_{-0.01}^{+0.02} \times 10^{-7}$&$4493.20/3602$\\
MAXI J1820+070&90401309006&58198.30&00010627010&58198.78&$1.52_{-0.00}^{+0.00}$&$18.54_{-0.25}^{+0.25}$&$5.35_{-0.05}^{+0.05}$&$1.33_{-0.01}^{+0.01} \times 10^{-7}$&$5195.07/3865$\\
MAXI J1820+070&90401309008&58201.52&00010627013&58201.11&$1.53_{-0.00}^{+0.00}$&$18.22_{-0.22}^{+0.31}$&$5.35_{-0.05}^{+0.05}$&$1.30_{-0.01}^{+0.01} \times 10^{-7}$&$4338.68/3608$\\
MAXI J1820+070&90401309010&58201.85&00010627014&58202.17&$1.53_{-0.00}^{+0.00}$&$18.43_{-0.28}^{+0.41}$&$5.30_{-0.06}^{+0.06}$&$1.34_{-0.02}^{+0.02} \times 10^{-7}$&$4476.90/3561$\\
MAXI J1820+070&90401309012&58212.19&00088657001&58212.20&$1.54_{-0.00}^{+0.00}$&$19.78_{-0.18}^{+0.18}$&$5.05_{-0.03}^{+0.03}$&$1.20_{-0.01}^{+0.01} \times 10^{-7}$&$6307.59/4206$\\
MAXI J1820+070&90401309013&58224.93&00010627043&58224.95&$1.58_{-0.00}^{+0.00}$&$22.75_{-0.80}^{+0.66}$&$4.40_{-0.09}^{+0.09}$&$1.15_{-0.01}^{+0.02} \times 10^{-7}$&$3871.94/3242$\\
MAXI J1820+070&90401309014&58225.27&00010627045&58225.47&$1.56_{-0.00}^{+0.00}$&$20.75_{-0.23}^{+0.23}$&$4.76_{-0.03}^{+0.03}$&$1.16_{-0.01}^{+0.01} \times 10^{-7}$&$6038.07/4112$\\
MAXI J1820+070&90401309016&58241.79&00088657003&58241.67&$1.59_{-0.00}^{+0.00}$&$23.18_{-0.29}^{+0.38}$&$4.27_{-0.04}^{+0.04}$&$1.01_{-0.01}^{+0.01} \times 10^{-7}$&$6766.12/4165$\\
MAXI J1820+070&90401309018&58255.15&00088657004&58255.90&$1.64_{-0.00}^{+0.00}$&$31.61_{-1.67}^{+1.58}$&$3.28_{-0.11}^{+0.11}$&$8.94_{-0.10}^{+0.09} \times 10^{-8}$&$4277.60/3270$\\
MAXI J1820+070&90401309019&58255.60&00088657004&58255.90&$1.63_{-0.00}^{+0.00}$&$30.32_{-1.01}^{+1.02}$&$3.42_{-0.07}^{+0.07}$&$8.98_{-0.09}^{+0.04} \times 10^{-8}$&$6821.45/3946$\\
MAXI J1820+070&90401309037&58404.95&00010627112&58404.26&$1.70_{-0.00}^{+0.01}$&$1000.00_{-567.61}^{+0.00}$&$0.19_{-0.05}^{+0.05}$&$2.24_{-0.08}^{+0.09} \times 10^{-9}$&$1975.57/1939$\\
MAXI J1820+070&90401309039&58420.05&00010627120&58419.60&$1.81_{-0.01}^{+0.00}$&$173.74_{-65.69}^{+191.84}$&$0.76_{-0.47}^{+0.47}$&$4.44_{-0.14}^{+0.17} \times 10^{-10}$&$2116.84/2065$\\
MAXI J1820+070&90501311002&58567.83&00010627149&58567.16&$1.69_{-0.00}^{+0.00}$&$148.69_{-60.93}^{+116.01}$&$1.01_{-0.48}^{+0.48}$&$2.02_{-0.06}^{+0.08} \times 10^{-9}$&$2283.13/2296$\\
MAXI J1820+070&90501320002&58604.40&\nodata&\nodata&$2.19_{-0.11}^{+0.18}$&$6.75_{-2.17}^{+2.66}$&$5.41_{-1.29}^{+1.29}$&$5.85_{-0.41}^{+0.64} \times 10^{-12}$&$871.47/998$\\
\hline
GRS 1915+105&90201053002&57840.67&00080228002&57840.82&$2.06_{-0.01}^{+0.01}$&$21.68_{-1.20}^{+1.40}$&$2.83_{-0.12}^{+0.12}$&$1.42_{-0.03}^{+0.03} \times 10^{-8}$&$3326.08/2846$\\
GRS 1915+105&90301001002&57875.18&00088091001&57875.48&$1.99_{-0.01}^{+0.01}$&$22.62_{-1.23}^{+1.63}$&$2.90_{-0.13}^{+0.13}$&$1.18_{-0.04}^{+0.03} \times 10^{-8}$&$3166.05/2830$\\
\hline
Cyg X-1&30001011005&56776.92&00080732001&56776.89&$1.70_{-0.00}^{+0.00}$&$27.75_{-1.07}^{+1.07}$&$3.37_{-0.09}^{+0.09}$&$3.06_{-0.04}^{+0.04} \times 10^{-8}$&$4258.02/3626$\\
Cyg X-1&30001011007&56797.24&00080732002&56798.01&$1.65_{-0.00}^{+0.00}$&$28.59_{-0.60}^{+0.78}$&$3.50_{-0.06}^{+0.06}$&$2.43_{-0.02}^{+0.02} \times 10^{-8}$&$5048.51/4163$\\
Cyg X-1&90101020002&57429.45&00081820001&57429.55&$1.60_{-0.00}^{+0.00}$&$30.04_{-0.91}^{+1.35}$&$3.59_{-0.09}^{+0.09}$&$2.67_{-0.02}^{+0.02} \times 10^{-8}$&$4346.34/3871$\\
Cyg X-1&30002150004&57537.89&00034310002&57537.54&$1.59_{-0.00}^{+0.00}$&$30.06_{-0.66}^{+0.53}$&$3.63_{-0.05}^{+0.05}$&$2.76_{-0.01}^{+0.01} \times 10^{-8}$&$6582.65/4512$\\
Cyg X-1&30002150006&57539.91&00034310003&57539.55&$1.59_{-0.00}^{+0.00}$&$30.16_{-0.64}^{+0.51}$&$3.62_{-0.04}^{+0.04}$&$2.63_{-0.01}^{+0.01} \times 10^{-8}$&$6364.94/4494$\\
Cyg X-1&30002150008&57541.92&00034310004&57541.48&$1.59_{-0.00}^{+0.00}$&$29.51_{-0.68}^{+0.75}$&$3.68_{-0.06}^{+0.06}$&$2.45_{-0.01}^{+0.01} \times 10^{-8}$&$5207.24/4264$\\
Cyg X-1&30202032002&57587.65&00081903001&57587.72&$1.59_{-0.00}^{+0.00}$&$27.39_{-0.84}^{+0.97}$&$3.84_{-0.08}^{+0.08}$&$2.51_{-0.02}^{+0.02} \times 10^{-8}$&$4016.08/3798$\\
\hline
V404 Cyg&30001010002&56578.50&00080264001&56579.78&$2.33_{-0.46}^{+1.01}$&$5.69_{-3.30}^{+8.85}$&$5.50_{-4.31}^{+4.31}$&$6.64_{-1.64}^{+2.53} \times 10^{-13}$&$127.48/154$\\
V404 Cyg&30001010003&56579.91&00080264001&56579.78&$2.08_{-0.13}^{+0.42}$&$6.48_{-3.04}^{+6.57}$&$5.98_{-2.90}^{+2.90}$&$7.38_{-1.47}^{+2.80} \times 10^{-13}$&$203.23/209$\\
V404 Cyg&30001010005&56628.73&00080264002&56628.75&$1.78_{-0.45}^{+0.60}$&$4.38_{-3.05}^{+4.24}$&$9.50_{-6.27}^{+6.27}$&$7.95_{-2.41}^{+7.63} \times 10^{-13}$&$79.74/79$\\
V404 Cyg&90102007005&57204.05&00031403068&57203.93&$1.72_{-0.03}^{+0.04}$&$17.02_{-2.00}^{+2.98}$&$4.47_{-0.44}^{+0.44}$&$2.19_{-0.14}^{+0.17} \times 10^{-10}$&$1148.85/1217$\\
V404 Cyg&90102007007&57209.28&00031403080&57209.51&$1.90_{-0.08}^{+0.08}$&$10.83_{-1.57}^{+2.76}$&$5.05_{-0.71}^{+0.71}$&$2.56_{-0.38}^{+1.74} \times 10^{-11}$&$363.57/449$\\
V404 Cyg&90102007009&57211.97&00031403085&57211.72&$1.88_{-0.03}^{+0.05}$&$16.10_{-2.92}^{+4.13}$&$4.00_{-0.58}^{+0.58}$&$3.10_{-0.43}^{+1.30} \times 10^{-11}$&$830.18/857$\\
V404 Cyg&90102007011&57226.35&00031403109&57226.40&$1.84_{-0.11}^{+0.19}$&$8.03_{-1.72}^{+4.69}$&$6.36_{-1.72}^{+1.72}$&$4.15_{-0.91}^{+1.49} \times 10^{-12}$&$220.02/249$\\
V404 Cyg&90102007013&57380.43&00031403122&57380.62&$1.81_{-0.02}^{+0.08}$&$39.44_{-3.73}^{+2.31}$&$2.35_{-0.18}^{+0.18}$&$8.45_{-2.34}^{+0.67} \times 10^{-11}$&$935.86/900$\\
V404 Cyg&90102007015&57393.64&00031403145&57393.57&$1.95_{-0.03}^{+0.09}$&$37.93_{-6.30}^{+4.74}$&$2.11_{-0.25}^{+0.25}$&$2.72_{-0.73}^{+0.71} \times 10^{-11}$&$690.02/751$\\
\hline
\hline
\enddata
\end{deluxetable*}
\end{longrotatetable}



\end{document}